# Single-Beam Magneto-Optical Trap in Back-to-Back Pyramidal and Conical Mirrors

Timothy H. Nguyen, Mariam Mchedlidze, Guanghui Su, Balthazar Loglia, Hanbo Yang[#], Xuejian Wu*

*Department of Physics, Rutgers University-Newark, Newark, NJ 07102, USA*
*[#]Present Address: Homer L. Dodge Department of Physics and Astronomy, The University of Oklahoma, Norman, OK, 73019, USA*
**Author to whom correspondence should be addressed: xuejian.wu@rutgers.edu*

**Abstract:** A three-dimensional magneto-optical trap (MOT), as an efficient method of producing cold atoms from room-temperature atomic vapor, has been widely used to develop atomic sensors. Various compact MOTs using a single laser beam have been reported, simplifying apparatuses and leading to miniaturized devices. Here, we propose single-beam MOTs based on back-to-back pyramidal and conical mirrors. In such back-to-back mirrors, a MOT trapping volume is formed by an incident laser beam, a retroreflected beam, and multiple reflections from the mirror surfaces. We present the design of back-to-back mirrors and a series of compact MOT configurations, with the potential of increasing access to the MOT and simultaneously creating multiple MOTs. We demonstrate a MOT in a back-to-back conical mirror, loading $\sim 1\times 10^7$ $^{87}$Rb atoms from background vapor and cooling the atoms to $\sim 7\ \mu$K using polarization gradients. Single-beam MOTs based on back-to-back mirrors will contribute to building compact and scalable cold-atom-based sensors.

## 1. Introduction

Trapping and cooling atoms have facilitated the development of many quantum sensors, such as atomic clocks for timekeeping [1, 2] and atom interferometers for inertial sensing [3, 4]. Magneto-optical traps (MOTs) [5], as the workhorse for creating a cold sample of neutral atoms, have been demonstrated in several configurations, from traditional setups requiring three orthogonal pairs of laser beams [6, 7] to compact ones using a single laser beam [8]. A typical MOT uses three pairs of counter-propagating red-detuned laser beams to apply radiation forces to the atoms based on the Doppler effect. Relying on laser beam reflections, single-beam MOTs have been demonstrated in pyramidal [9-14], conical [15, 16], and tetrahedral [17] mirrors. Additionally, single-beam MOTs have been achieved using diffractive optics, such as gratings [18-22], Fresnel reflectors [23, 24], and metasurface optical chips [25]. Such single-beam MOTs simplify the complexity of preparing cold atoms and give rise to transportable atomic gravimeters [26, 27] and gradiometers [28, 29], compact atomic inertial sensors [30, 31], and miniaturized atomic clocks [32, 33].

The development of cold-atom-based sensors will benefit from novel MOT configurations. A single-beam MOT that can simultaneously create multiple cold atomic clouds will reduce the complexity of multiaxis atom interferometers [34, 35]. Additionally, compact MOTs with rich access to light and particle beams will enable interactions with the cold atoms, contributing to the generation of cold electron beams [36] and the search for massive neutrinos [37].

Here, we propose and demonstrate single-beam MOTs based on back-to-back pyramidal and conical mirrors. As a modification of pyramidal MOTs [9-12], Pollock proposed a MOT geometry by placing two pyramidal mirrors with a top angle of 72° back-to-back and opening a central through-hole [13]. We describe a general design of back-to-back mirrors and bring about a series of single-beam MOT configurations. Using a back-to-back conical mirror, we load millions of $^{87}$Rb atoms from the background vapor and cool the atoms to a few microkelvins using polarization gradients. Single-beam MOTs based on back-to-back mirrors provide a compact solution to create cold atomic clouds, with the potential of introducing access to the MOT in a miniaturized setup and creating multiple MOTs using a single laser beam.

## 2. MOT in Back-to-Back Mirror

Figure 1 (a) shows the MOT schematic based on a back-to-back mirror. As a laser beam illuminates the back-to-back mirror, the center rays of the incident laser beam pass directly through the hole. The back-to-back mirror reflects the side rays twice, flipping from an inner position to an outer position and vice versa. Thus, the incident laser beam passes through the back-to-back mirror and remains collimated. When a retroreflector, consisting of a quarter-wave plate and a flat mirror, reflects the laser beam to its original path, a trapping volume is formed in the center of the back-to-back mirror. In a 2D schematic, the trapping area is a diamond shape with six overlapped laser beams.

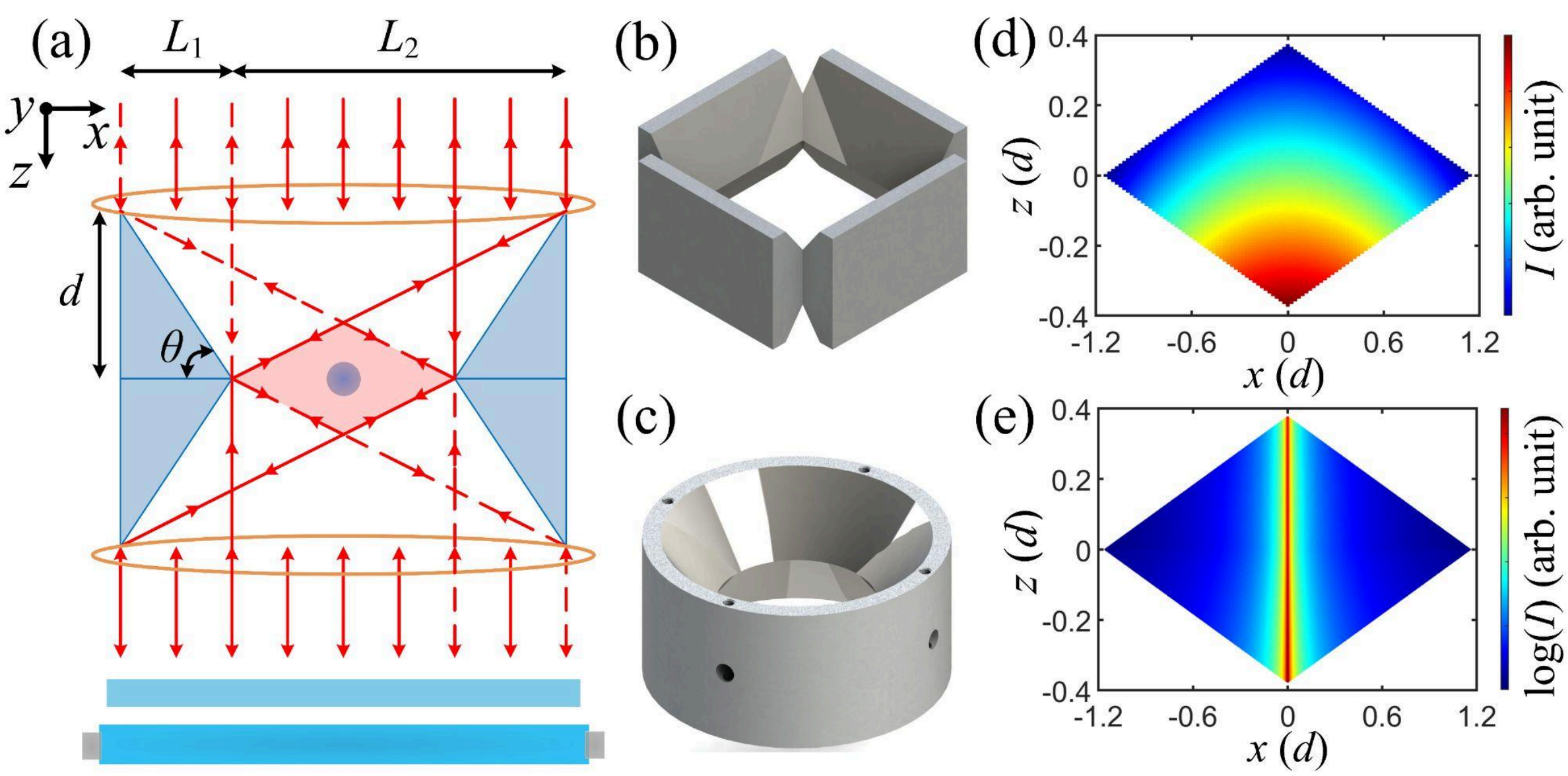

FIG. 1. MOT in a back-to-back mirror. (a) Cross-sectional MOT schematic. The blue triangles represent the prisms of the back-to-back mirror. The orange ellipses represent the anti-Helmholtz coils. The blue rectangles under the prisms are a quarter-wave plate and a flat mirror. The red lines and arrows represent the laser beams. The dashed lines highlight the typical optical paths. The red diamond-shaped area depicts the MOT trapping area. The blue sphere represents a cold atomic cloud. (b) 3D model of a back-to-back pyramidal mirror. (c) 3D model of a back-to-back conical mirror. (d) Simulated laser intensity distribution of the 2D trapping area in a back-to-back pyramidal mirror. (e) Simulated laser intensity distribution of the 2D trapping area in a back-to-back conical mirror.

The reflection surface of a back-to-back mirror can be flat or conical, forming the so-called back-to-back pyramidal or conical mirror, as shown in Fig. 1 (b) and (c), respectively. The trapping volume of a back-to-back pyramidal mirror is a double pyramid, and that of a back-to-back conical mirror is a double cone. Assuming the incident laser beam has a Gaussian profile, Fig. 1 (d) and (e) show the simulated laser intensity distributions in the respective 2D trapping areas. In a back-to-back pyramidal mirror, the rays closer to the center of the incident laser beam are reflected toward the lower region of the trapping volume. Thus, the laser intensity distribution has a gradient along the $z$-axis. In a back-to-back conical mirror, the incident laser beam is focused on a line in the center of the trapping volume, resulting in a strong laser intensity gradient along the radial axis. The laser intensity distribution of the trapping volume in back-to-back pyramidal and conical mirrors is similar to that in traditional pyramidal and conical mirrors [9, 15].

To optimize the dimensions and inclination angle of a back-to-back mirror, we define the half-height of the back-to-back mirror as $d$ and the inclination angle of the reflection surface as $\theta$, as shown in Fig. 1 (a). We assume that the incident rays reflected by the outer edge of the mirror surface arrive precisely at the inner edge. The length of the back-to-back mirror along the $x$-axis is $L_1+L_2$, and the diameter of the central hole is $L_2$-$L_1$, where $L_1$=$d$/tan$\theta$ and $L_2$=-$d$·tan(2$\theta$). The angle between the reflected laser beams and the $z$-axis is 180°-2$\theta$. Figure 2 shows the back-to-back mirror dimensions and trapping volumes as a function of $\theta$. Ideally, $\theta$ can be any angle between 45° and 90°. A smaller $\theta$ leads to a wider back-to-back mirror and thus a larger trapping volume.

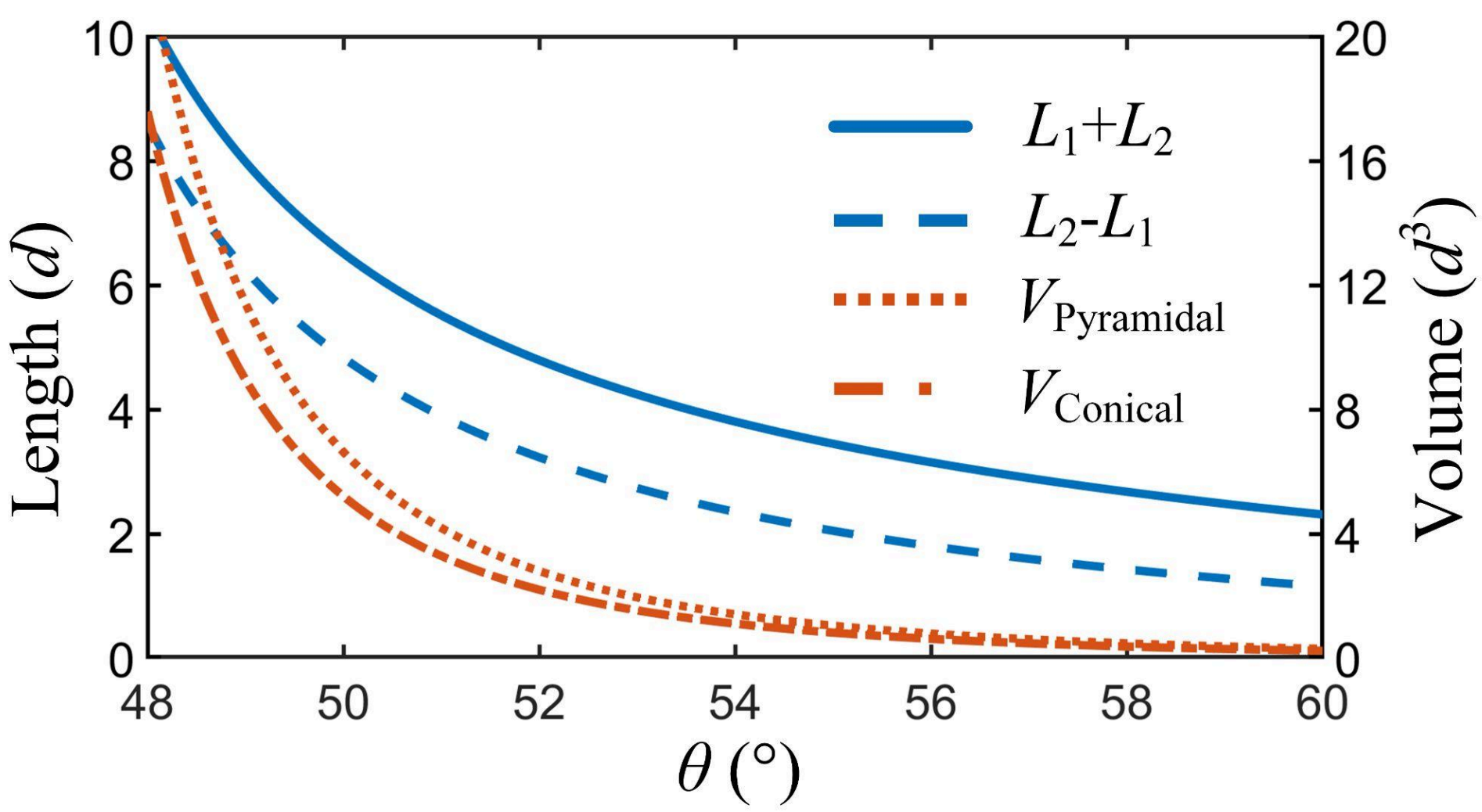

FIG. 2. Back-to-back mirror dimensions and trapping volumes as a function of $\theta$.

## 3. Single-beam MOT Configurations

We propose two single-beam MOT configurations to introduce access to the cold atomic cloud. Figure 3 (a) shows that a hollow-core laser beam can illuminate the back-to-back mirror to create a MOT. Though the rays passing through the central hole of the back-to-back mirror are blocked, the reflection beams from the back-to-back mirror produce the MOT trapping volume. We can use the center space of the incident laser beam to send a probing laser beam or detect the fluorescence of the atomic cloud. Figure 3

(b) shows a MOT configuration with a gap separated by the upper and lower sets of the back-to-back mirror. As the separation distance increases, the MOT trapping volume will decrease. The gap between the upper and lower sets of the back-to-back mirror can provide additional access to the MOT. Note that these two single-beam MOT configurations can be combined to allow laser and particle beams along all axes to reach the MOT.

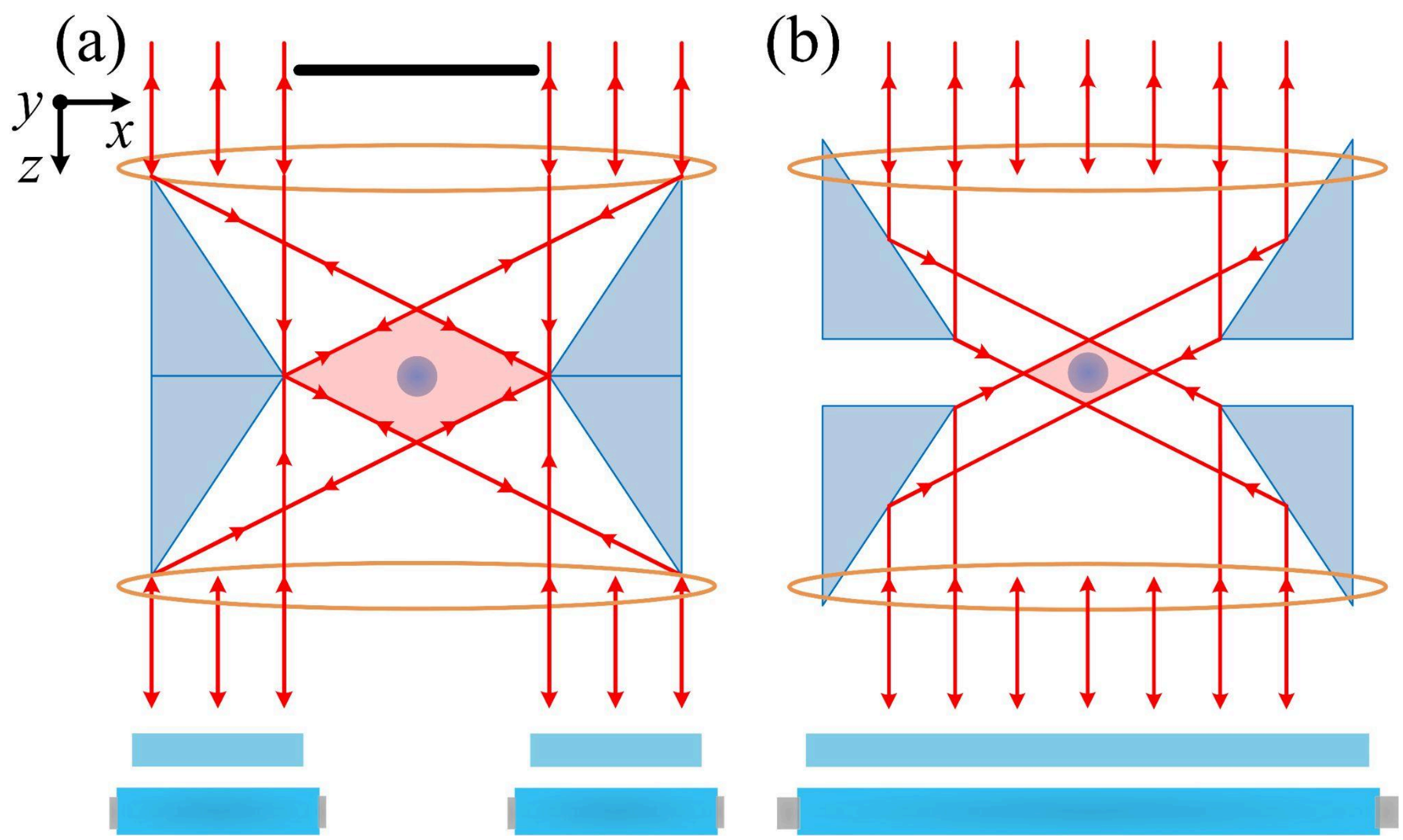

FIG. 3. MOT configurations with access to the atomic cloud. (a) MOT with a hollow-core incident laser beam. (b) MOT with a gap between the upper and lower sets of the back-to-back mirror.

Additionally, we propose a scalable MOT configuration by stacking back-to-back mirrors. Figure 4 (a) shows the schematic of a single laser beam illuminating two aligned back-to-back mirrors to create two MOTs simultaneously. Because the laser beam passing through the back-to-back mirrors remains collimated, the trapping volume can be formed in each back-to-back mirror.

To verify the feasibility, we mount two silver-coated aluminum back-to-back conical mirrors on a 3D-printed spacer (see Fig. 4 (b) and (c)), illuminate them with a collimated laser beam, and observe the reflected patterns on a viewing screen. As shown in Fig. 4 (e) and (f), the laser beam passing through the back-to-back conical mirror has an inner circle and an outer ring. The inner circle is the laser beam passing through the central hole, and the outer ring is the laser beam reflected by the back-to-back conical mirror. Because the back-to-back conical mirrors have been exposed to air for a long time, the surface has deteriorated reflectivity, causing an inhomogeneous ring pattern. Due to imperfect reflection on the edge of the conical surface, a small ring-shaped gap exists between the inner circle and the outer ring. Since the rays reflected on the conical surface's edge contribute to the trapping volume's outer layer, the MOT is unaffected by the imperfect edge reflections. Comparing the incident and reflected patterns, the laser beam after the back-to-back conical mirrors remains collimated, and the diffraction effect of the back-to-back mirrors is negligible at a distance of ~0.3 m.

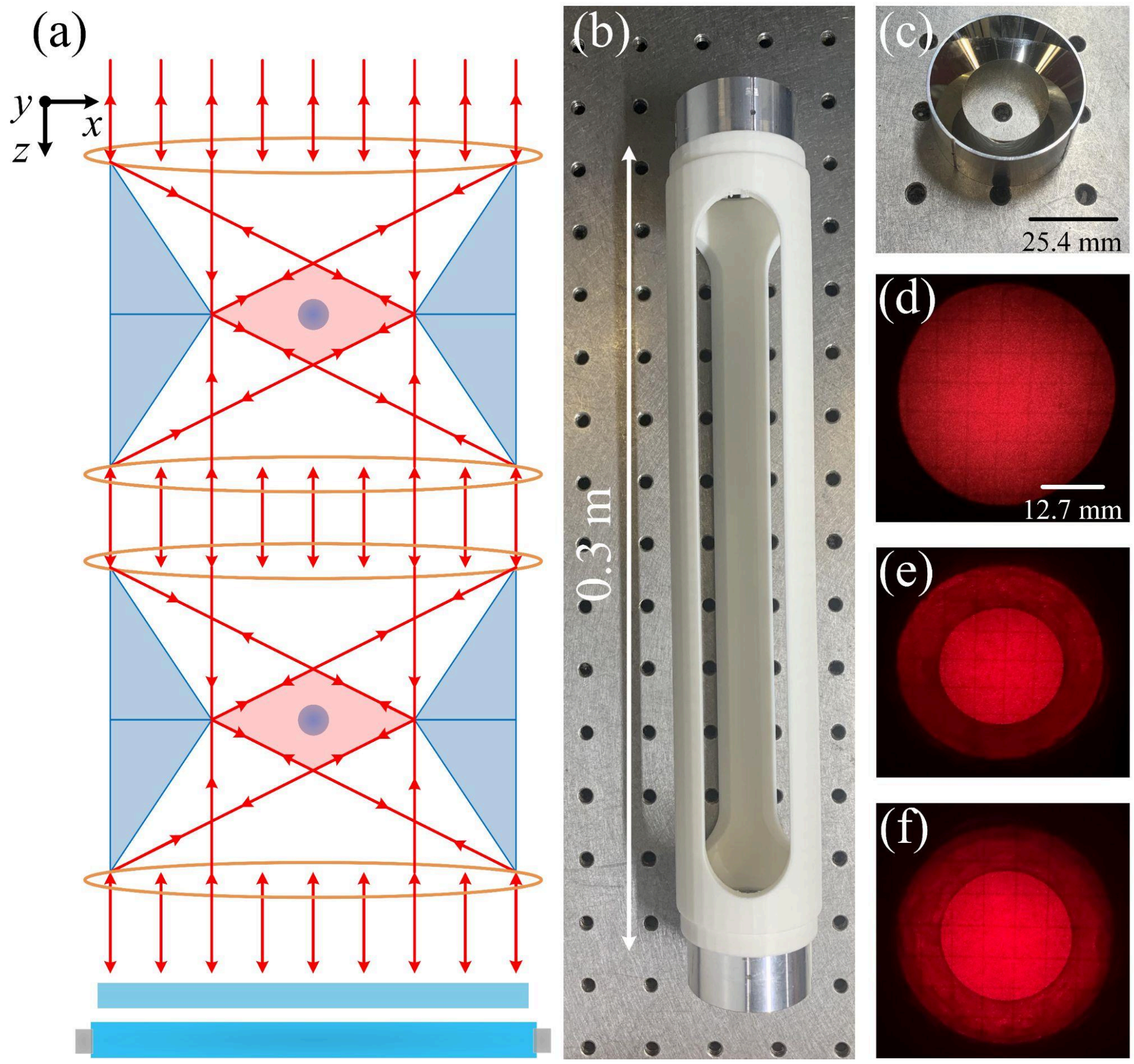

FIG. 4. Scalable MOT configuration. (a) Schematic of two simultaneous MOTs. (b) Picture of two back-to-back conical mirrors mounted on a 3D-printed spacer. (c) Picture of a back-to-back conical mirror. (d) Laser beam before the first back-to-back conical mirror. (e) Laser beam after the first back-to-back conical mirror observed at a distance of ~0.3 m. (f) Laser beam after the second back-to-back conical mirror observed at a distance of ~0.3 m.

**4. Experimental Setup**

We use a back-to-back conical mirror to demonstrate the MOT. As shown in Fig. 4 (c), the back-to-back conical mirror is made of aluminum by a computer numerical control machine. The conical surfaces are polished to a root-mean-squared surface deviation of ~120 nm and coated with protected silver. The reflectivity of the conical surface is about 95% at 780 nm. We choose metallic coatings to preserve circular polarization upon reflection. Dielectric coatings can be used as well if an equal phase shift between the two orthogonal polarizations can be achieved. The half-height of the mirror is 12.4 mm. The inclination angle of the reflection surface is 54.7°. The outer diameter of the mirror is 44 mm, and the diameter of the central hole is 26.4 mm. The MOT trapping volume is 1.7 $cm^3$.

As shown in Fig. 5, our experimental setup is based on a 3.38-inch cubic vacuum chamber (CU6-0337, Kurt J. Lesker). The back-to-back conical mirror is held in the center of the vacuum chamber by four aluminum rods. A Rb dispenser (RB/NF/4.8/17 FT10+10, SAES) is mounted on an electrical feedthrough at one side of the vacuum chamber. A 5 L/s ion pump (5S TiTan, Gamma) keeps the background pressure ~$1.5\times10^{-9}$ Torr. The magnetic field is controlled by an anti-Helmholtz coil pair installed outside the top and bottom viewports and three Helmholtz coil pairs around the vacuum chamber. The magnetic gradient around the MOT trapping volume is 12 G/cm. The incident laser beam illuminates the back-to-back conical mirror from the top viewport. A retroreflector consisting of a quarter-wave plate and a flat mirror under the bottom viewport reflects the laser beam back to its original path. The transmission loss through the viewport is about 5%. A digital camera is positioned towards the top viewport to observe the MOT, and another digital camera is positioned towards a side viewport to image the freely falling atomic cloud.

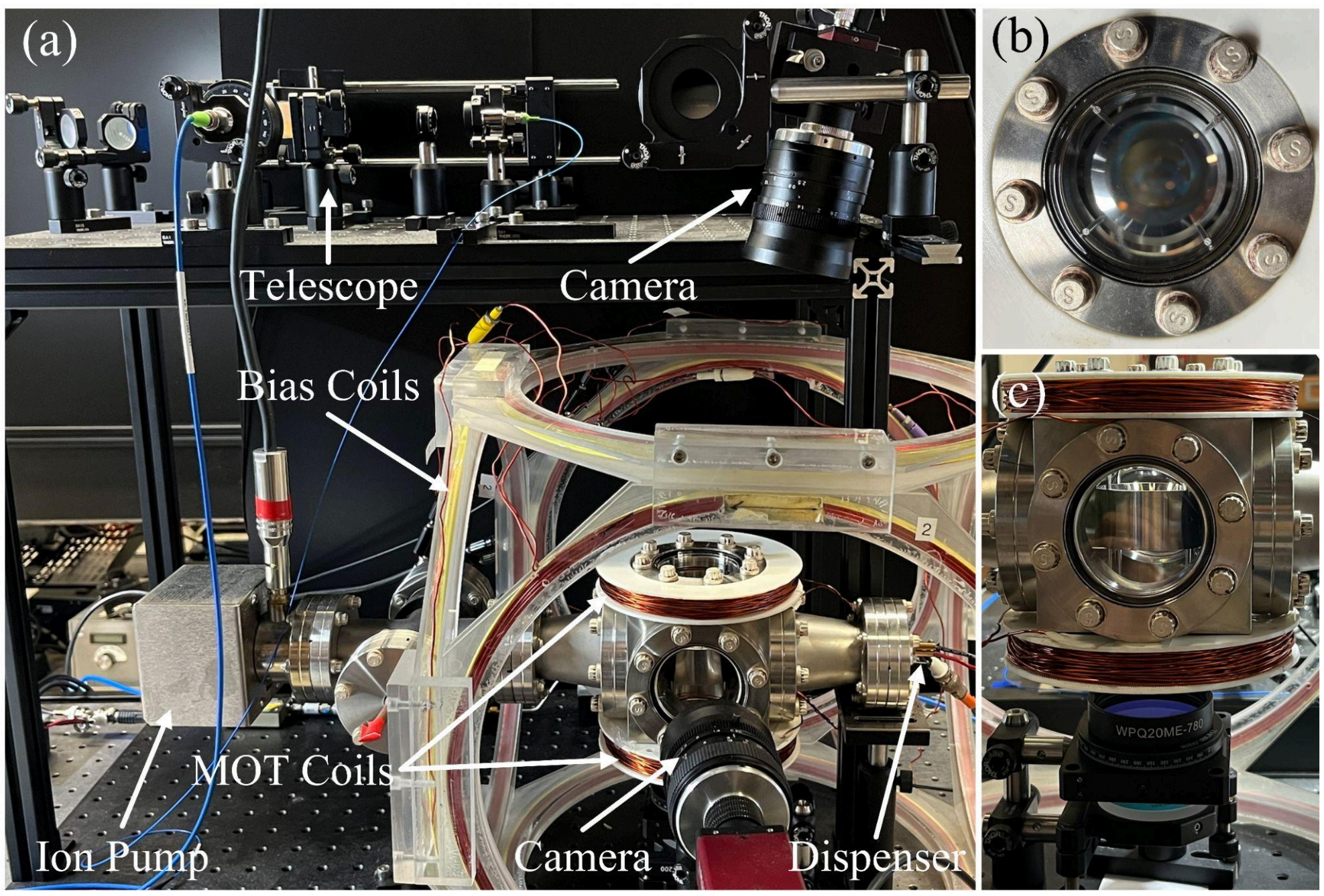

FIG. 5. Experimental apparatus. (a) Vacuum chamber and optics. (b) Top view of the vacuum chamber. (c) Side view of the vacuum chamber.

The laser system is based on modulating a 780-nm, 180-mW distributed Bragg reflector laser (780.241DBRH-MHFL-TO8, Photodigm) [27, 30]. The main laser output serves as the cooling beam. To adjust the frequency detuning of the cooling beam, the laser is modulated by a fiber-based electro-optical modulator (EOM), and the +1st order sideband is locked to the $^{85}$Rb $F$=3 to $F'$=4 transition via modulation transfer spectroscopy [38]. A diffracted part of the laser output modulated by another fiber-based EOM is the repumping beam, resonating at the $^{87}$Rb $F$=1 to $F'$=2 transition. The cooling beam with a laser power of ~27 mW and the repumping beam with ~1 mW are combined at a polarizing beam splitter and then collimated to a diameter of ~54 mm ($1/e^2$). When loading atoms from the background vapor to the MOT, the cooling beam is detuned ~7 MHz below the $^{87}$Rb $F$=2 to $F'$=3 transition. After that, a polarization gradient cools the atoms further by gradually increasing the detuning and reducing the power of the cooling beam. As the atomic cloud freely falls out of the MOT trapping volume, the two vertically counter-propagating beams passing directly through the central hole of the back-to-back conical mirror serve as the detection beam. The detection beam consists of a cooling frequency resonating at the $^{87}$Rb $F$=2 to $F'$=3 transition and the repumping frequency resonating at the $^{87}$Rb $F$=1 to $F'$=2 transition.

## 5. MOT Characterization

Figure 6 (a) shows MOT fluorescence images as we verify the laser alignment and the MOT quadrupole magnetic field. We observe ball- and ring-shaped MOTs similar to traditional conical MOTs [15, 16]. The shape of the MOT varies with the zero-magnitude position of the quadrupole magnetic field and the alignment between the counter-propagating laser beams and the back-to-back conical mirror. The wave vectors of the reflected laser beams are not parallel to the lines of the quadrupole magnetic field, introducing unwanted polarization components. Additionally, the conical reflection surfaces focus the laser beams to a line on the $z$-axis in the center of the trapping volume, such that the laser alignment changes the laser intensity distribution in the trapping volume. Thus, the MOT approaches a ring shape as the zero magnitude position of the quadrupole magnetic field moves toward the center of the trapping volume, and the counter-propagating laser beams perpendicularly align with the back-to-back conical mirror.

We observe a ring-shaped MOT, a dual-MOT, and a tri-MOT shown in Fig. 6 (b) when we misalign the retroreflected laser beam between 2° and 3°. The top and bottom MOT coils are at 3 A. The incident laser beam is perpendicular to the back-to-back conical mirror. A misaligned reflected laser beam modifies the laser intensity distribution in the MOT trapping volume and thus changes the MOT shapes.

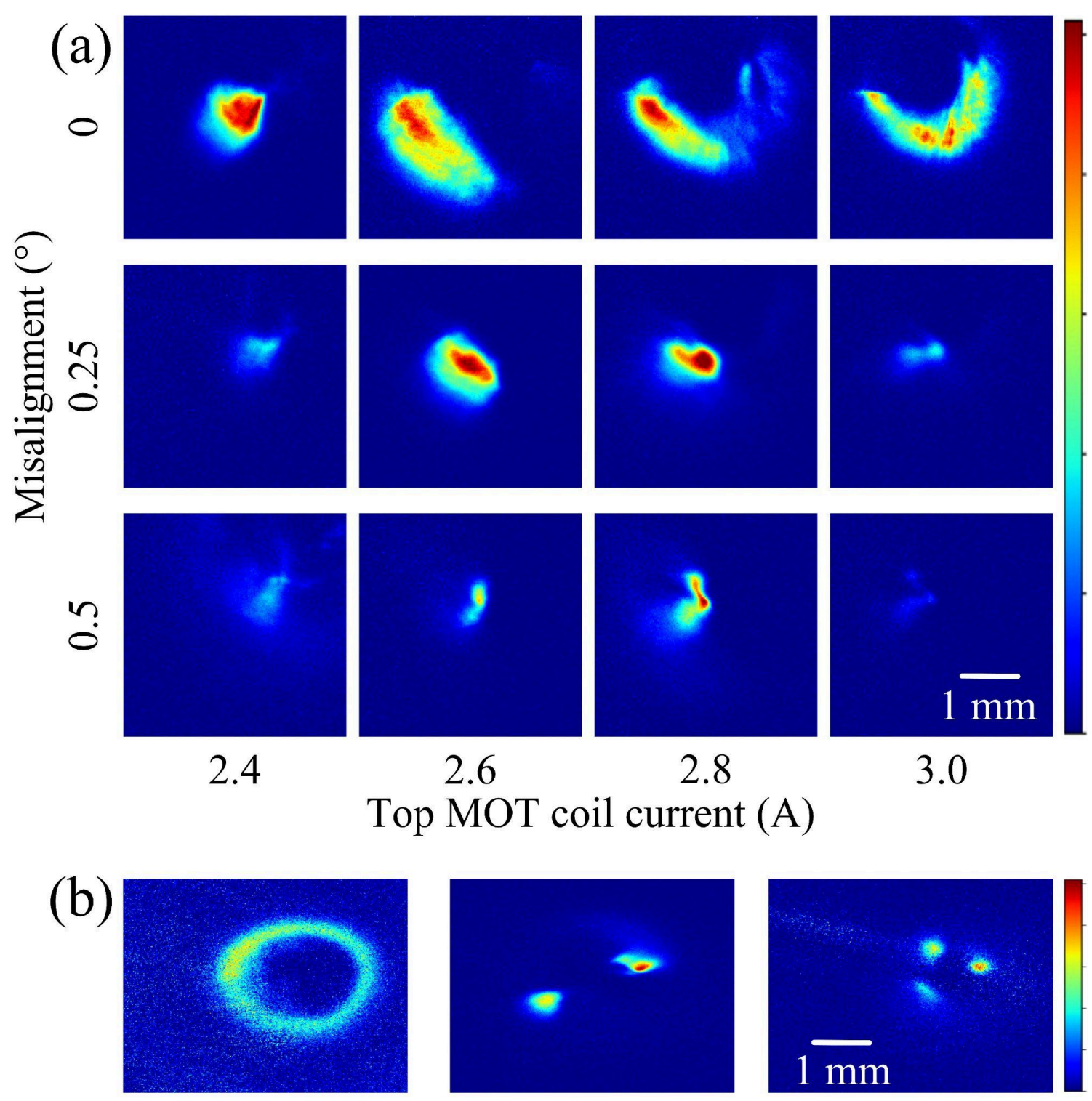

FIG. 6. MOT fluorescence images. (a) MOT variations as a function of misalignment and quadrupole magnetic field. The misalignment indicates the angle between the vertical axis of the back-to-back conical mirror and the axis of the counter-propagating laser beams. The bottom MOT coil is constant at 3 A. Each MOT coil has a diameter of ~10 cm and contains 120 turns of copper wire. (b) Ring-shaped MOT (Left), dual-MOT (Middle), and tri-MOT (Right).

Additionally, we obtain MOTs as we partially block the incident laser beam by inserting a circular object with a 10-mm diameter in the optical path (see Fig. 7 (a)), demonstrating the proposed MOT configuration shown in Fig. 3 (a). The object blocks a part of the laser beam passing through the center hole of the back-to-back conical mirror. As we move the object around, the MOT is tolerant of the object's position. As shown in Fig. 7 (e) and (f), the atom number and the shape of the MOT are similar to the original MOT without blocking the incident laser beam.

Furthermore, to verify the scalable MOT configuration shown in Fig. 4 (a), we place another back-to-back conical mirror above the vacuum chamber viewport and position it with the one in the vacuum chamber (see Fig. 7 (b)). The distance between the two back-to-back conical mirrors is ~16 cm. As shown in Fig. 7 (g) and (h), the shape and position of the MOT are changed due to the tilt between the two back-to-back conical mirrors, and the atom number in the MOT decreases due to the laser power loss through the above back-to-back conical mirror.

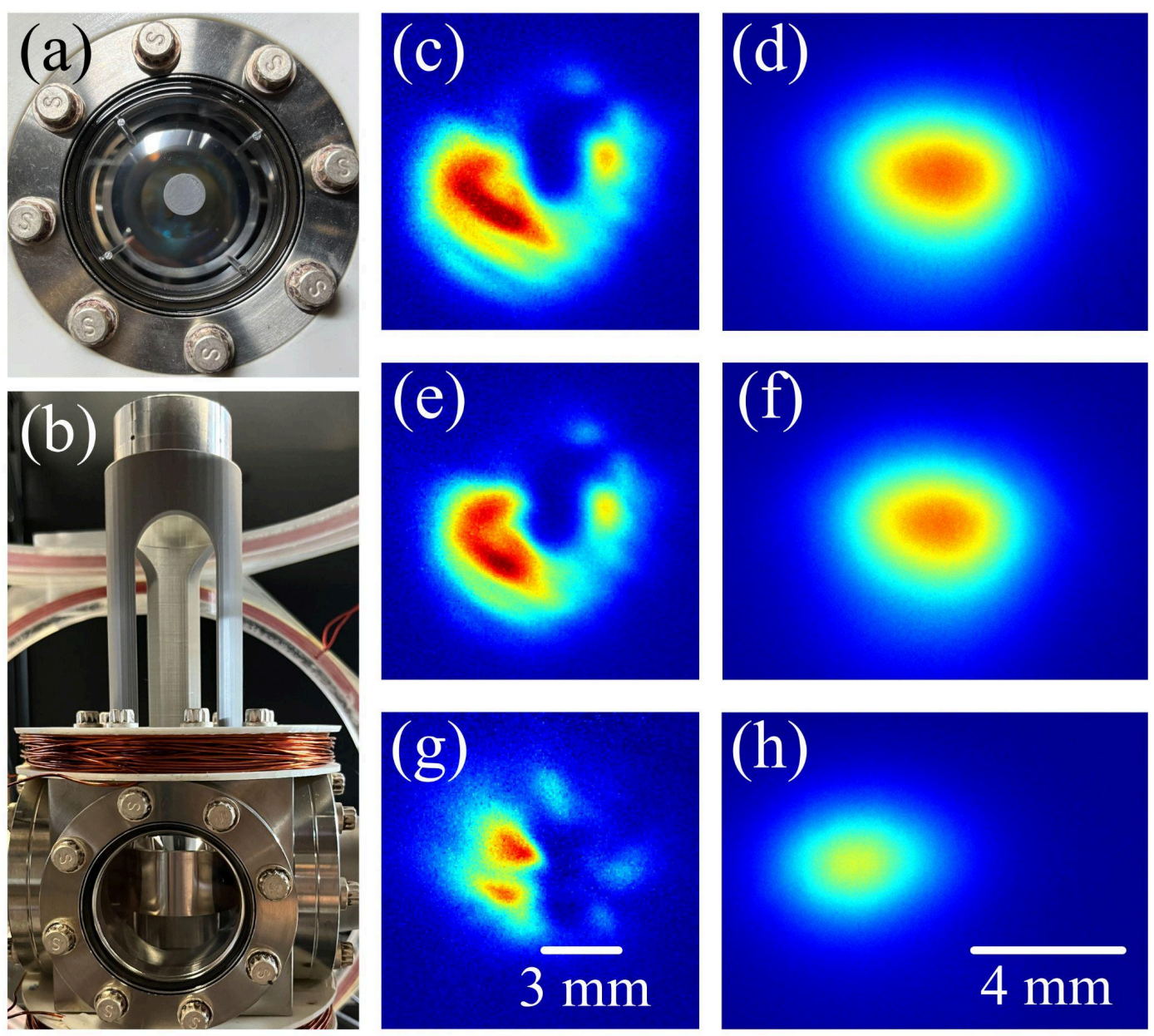

FIG. 7. MOT fluorescence images with blocking and stacking the back-to-back conical mirror. (a) The incident laser beam is partially blocked by a 10-mm-diameter object. (b) Stacking another back-to-back conical mirror on top of the vacuum chamber. (c) Original MOT. (d) Original MOT after 70-ms free-fall. (e) Blocking MOT. (f) Blocking MOT after 70-ms free-fall. (g) Stacking MOT. (h) Stacking MOT after 70-ms free-fall.

We optimize the MOT by adjusting the laser alignment and the quadrupole magnetic field. By observing via the camera towards the top viewport, we first adjust the MOT close to a ball shape. Then, we make fine adjustments by observing the freely falling atomic cloud via fluorescence imaging. After polarization gradient cooling, all the MOTs shown in Fig. 6 become a Gaussian atomic cloud and reach a similar temperature. Due to intensity imbalance caused by the loss of the retroreflector as well as the Gaussian beam profile of the incident laser beam, we obtain the maximum atom number when the zero-magnitude position of the quadrupole magnetic field is above the center of the MOT trapping volume. Besides, we obtain the maximum atom number when the misalignment between the laser beams and the back-to-back conical mirror is close to zero, maximizing the MOT trapping volume. For example, comparing the MOTs shown in Fig. 6 (a), the one with a misalignment angle of 0° and a top MOT coil current of 2.6 A contains the maximum atom number.

We measure the temperature of the atomic cloud using the time-of-flight method, where the width of the atomic cloud is measured as a function of free-fall time. The temperature of the atomic cloud $T$ is related to its full width at half maximum $\sigma(t)$, expressed as $\sigma(t)=(\sigma_0^2+k_B T t^2/m)^{1/2}$, where $\sigma_0$ is the initial width of the atomic cloud, $k_B$ is the Boltzmann constant, $t$ is the free-fall time, and $m$ is the atomic mass. $\sigma(t)$ is measured by imaging the free-fall atomic cloud and fitting the spatial atom population distribution using a Gaussian function. Figure 8 (a) shows a free-fall atomic cloud with a temperature of ~7 $\mu$K. The size of the vacuum chamber limits the free-fall time.

We measure the MOT loading rate with different Rb vapor pressures. The Rb vapor pressure is varied by adjusting the dispenser current and is monitored by the ion pump. The atom number is calculated by imaging the atomic cloud with a 65-ms free-fall time. Figure 8 (b) shows the measurement results. The data is fit to an exponential growth function $N(t)=N_0(1-e^{-t/\tau})$, where $N(t)$ is the atom number at the MOT loading time $t$, $N_0$ is the steady-state atom number, and $\tau$ is the time constant. The saturated atom number in the free-fall atomic cloud is ~$1\times10^7$.

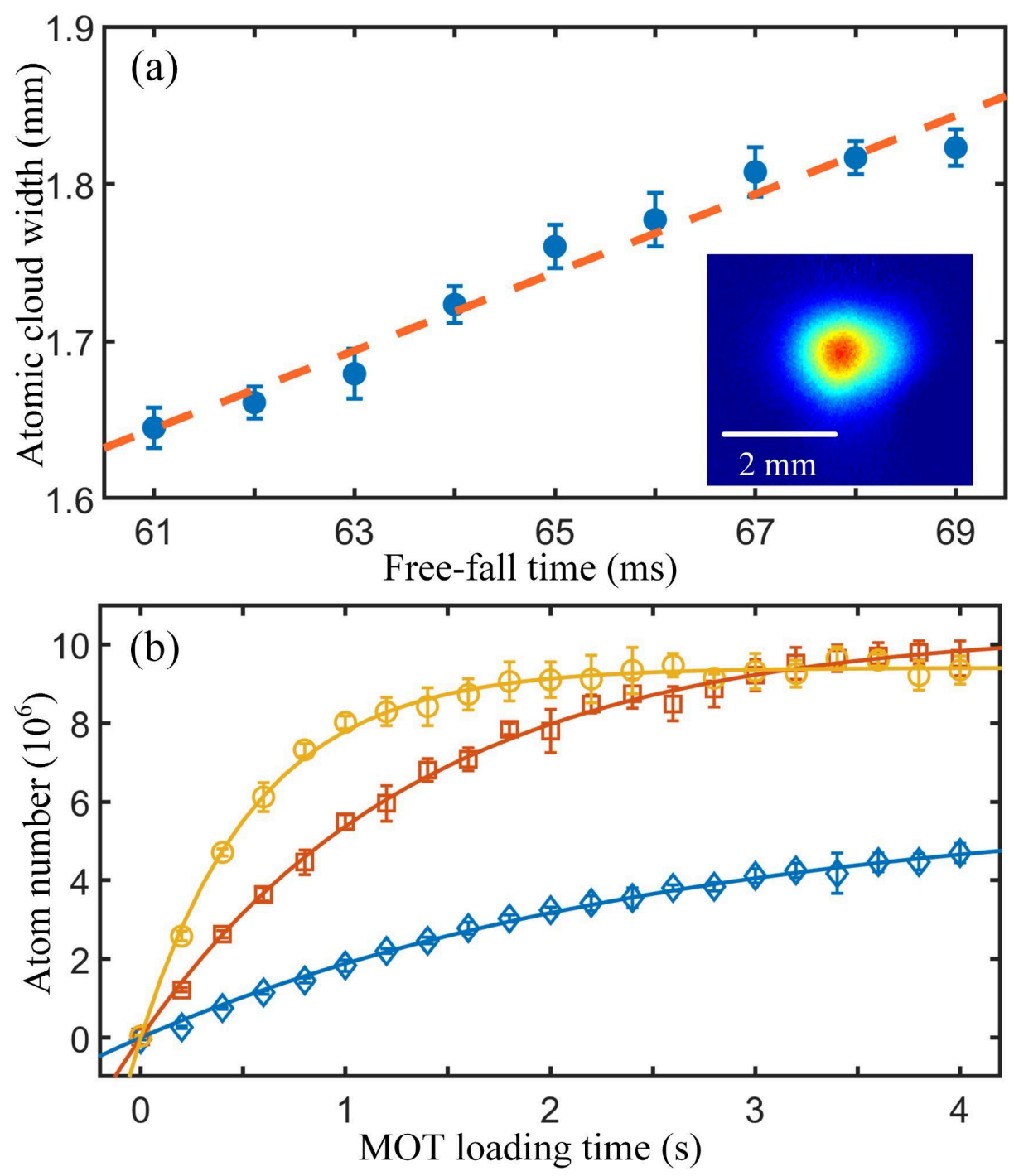

FIG. 8. (a) Temperature measurement. Each data point is averaged from ten measurements, and the error bars are 1-$\sigma$ statistical errors. The dashed curve is a fit using the time-of-flight method. The inserted picture is an image of the atomic cloud at a free-fall time of 65 ms. (b) Atom number measurement. Each data point is averaged from five measurements, and the error bars are 1-$\sigma$ statistical errors. The solid curves are fits based on the exponential growth function. The Rb vapor pressures of the blue diamonds, orange squares, and yellow circles are $2.7\times10^{-9}$, $3.5\times10^{-9}$, and $7.5\times10^{-9}$ Torr, respectively.

Figure 9 shows the dependence of the MOT atom number on the incident cooling laser intensity. Compared to a traditional pyramidal mirror MOT with a similar trapping volume, the atom number of the back-to-back conical mirror MOT saturates more quickly as the cooling laser intensity increases. The reason is that the conical surfaces focus the cooling laser in the center of the trapping volume, increasing the cooling beam intensity. Similarly, a lower cooling beam intensity in generating cold atomic beams was reported using a conical mirror funnel [39] compared to a pyramidal mirror funnel [14].

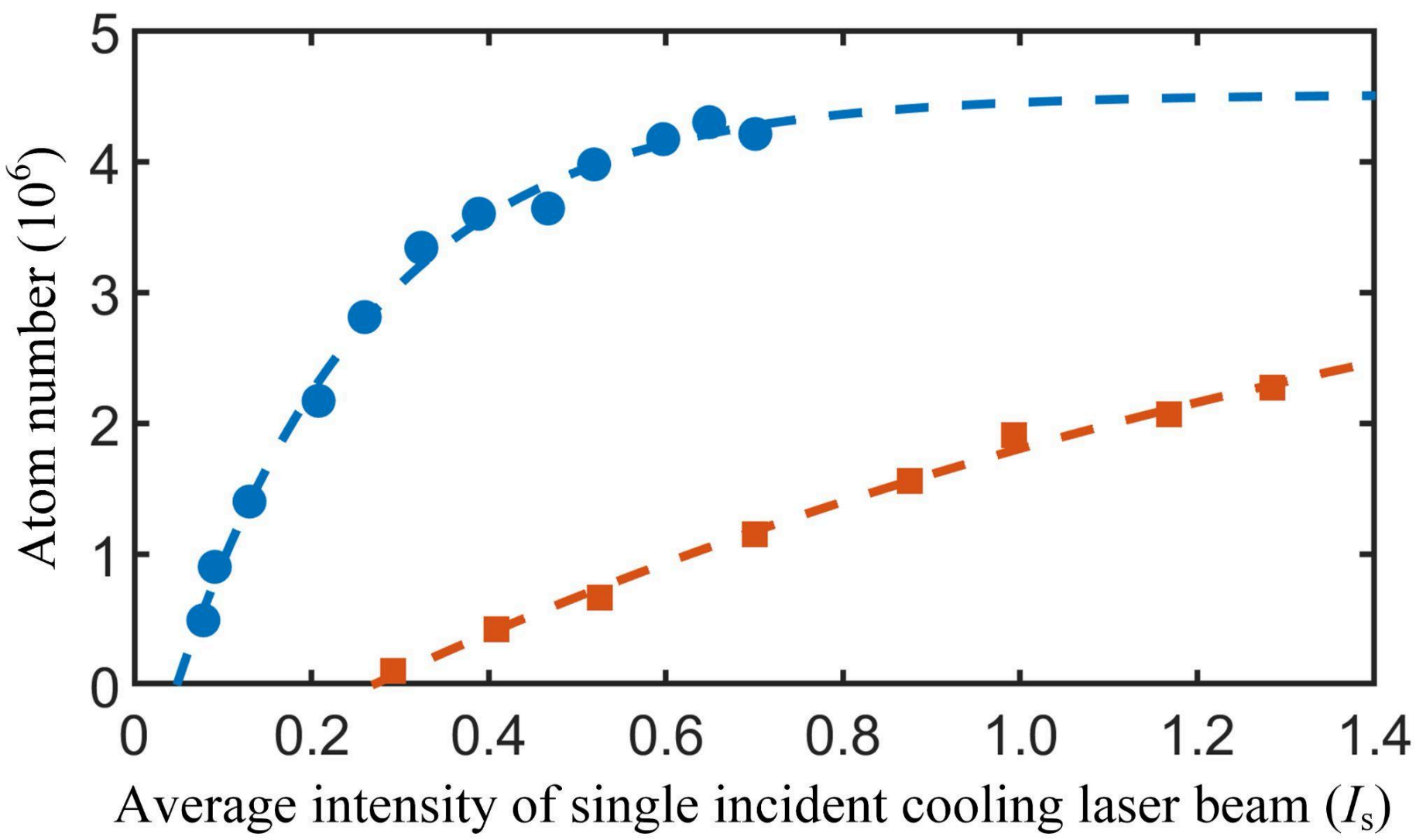

FIG. 9. MOT atom number as a function of the average intensity of a single incident cooling laser beam. The blue circles are measurements of the MOT in a back-to-back conical mirror, with a laser beam diameter of ~5.4 cm ($1/e^2$). The orange squares are measurements of a MOT in a traditional pyramidal mirror, with a laser beam diameter of ~3.6 cm ($1/e^2$). Each measurement is averaged from five times, and the error bars are smaller than the dots. The dashed curves are exponential fits. The saturation intensity $I_s$ of the $^{87}$Rb $F$=2 to $F'$=3 transition with $\sigma^{\pm}$-polarized light is 1.67 mW/cm$^2$.

We investigate the long-term MOT stability by measuring the atom number and temperature fluctuation over 36 hours. The Rb vapor pressure is $2.5\times10^{-9}$ Torr, and the MOT loading time is 1.5 s. The atom number and temperature are simultaneously measured by imaging the free-fall atomic cloud. The average atom number is $1.5\times10^6$. Assuming the atomic clouds have a constant initial width, the average temperature of the atomic clouds is 7.4 $\mu$K. The peak-to-peak fluctuations of the atom number and temperature are about 20%, mainly caused by the laser power and polarization drifts of the cooling beam.

**6. Conclusion**

We propose single-beam MOTs based on back-to-back mirrors and demonstrate a MOT in a back-to-back conical mirror. We verify the dependence of the MOT on the laser alignment and quadrupole magnetic field, observing ball- and ring-shaped MOTs inside the double-cone trapping volume. We load ~$1\times10^7$ $^{87}$Rb atoms from the background vapor and cool the atoms to ~7 $\mu$K using polarization gradient cooling. The atom number and the long-term stability of the MOT can be further improved by a cooling beam with higher intensity and lower power drifts.

Back-to-back mirrors offer an alternative way to create MOTs using a single laser beam. We illustrate MOT configurations to increase access to the MOT by blocking the center part of the incident laser beam and adding a gap between the upper and lower sets of the back-to-back mirrors. These MOT configurations will maintain the compact feature of using a single laser beam and provide flexibility to interact with the cold atomic cloud. Additionally, back-to-back mirrors preserve a collimated pass-through laser beam. This feature may allow the creation of two MOTs using a single laser beam to illuminate two identical sets of spatially separated back-to-back mirrors, enabling the development of compact atomic gradiometers.

**ACKNOWLEDGEMENT**

We thank Jose Dominguez, Maurice Metivier, Aditya Raman, Juan Chang, Shourya Chhabra, and Karina Ortiz for contributing to the experiment. We also thank Holger Müller, Peter Stromberger, and Storm Weiner for discussions. This work is supported by the National Science Foundation under award numbers 2328663 and 2316595, the Rutgers Research Council Award, and supplement funding provided by the Rutgers University–Newark Chancellor's Research Office.

**AUTHOR DECLARATIONS**

**Conflict of Interest**

T. H. Nguyen, M. Mchedlidze, G. Su, and X. Wu have a pending patent application.

**DATA AVAILABILITY**

The data supporting the findings of this study is available from the corresponding author upon reasonable request.